\documentclass[a4paper]{article}

\usepackage{INTERSPEECH2022}

\usepackage{hyperref}
\usepackage{booktabs}

\title{Neural Lexicon Reader: Reduce Pronunciation Errors in End-to-end TTS by Leveraging External Textual Knowledge}
\name{Mutian He \thanks{$^{\star}$ Work done at Microsoft. Corresponding author: Lei He.} $^{1 \star }~~~~~~~~$
  Jingzhou Yang$^2~~~~~~~~$
  Lei He$^2~~~~~~~~$
  Frank K. Soong$^2$ }
\address{
  $^1$ 
  The Hong Kong University of Science and Technology 
  $^2$
  Microsoft China
}

\email{
  mhear@connect.ust.hk,
  \{jingy,helei,frankkps\}@microsoft.com
}

\begin{document}

\maketitle
\begin{abstract}
End-to-end TTS requires a large amount of speech/text paired data to cover all necessary knowledge, particularly how to pronounce different words in diverse contexts, so that a neural model may learn such knowledge accordingly. But in real applications, such high demand of training data is hard to be satisfied and additional knowledge often needs to be injected manually. For example, to capture pronunciation knowledge on languages without regular orthography, a complicated grapheme-to-phoneme pipeline needs to be built based on a large structured pronunciation lexicon, leading to extra, sometimes high, costs to extend neural TTS to such languages. In this paper, we propose a framework to learn to automatically extract knowledge from unstructured external resources using a novel Token2Knowledge attention module. The framework is applied to build a TTS model named Neural Lexicon Reader that extracts pronunciations from raw lexicon texts in an end-to-end manner. Experiments show the proposed model significantly reduces pronunciation errors in low-resource, end-to-end Chinese TTS, and the lexicon-reading capability can be transferred to other languages with a smaller amount of data.%, which supports the potential of our framework.
\end{abstract}

\section{Introduction}
\label{sec:intro}

End-to-end neural models have greatly pushed the frontiers of the area of text-to-speech (TTS) \cite{DBLP:conf/icassp/ShenPWSJYCZWRSA18}. However, these models still suffer from high data demands.
As a machine learning method, a large amount of data is generally required to cover all knowledge necessary to produce speech: pronunciation of each grapheme under different contexts, semantic meanings of words to produce natural prosody, etc. The cost to collect such a complete dataset can be unbearable, and knowledge learned from limited data may fail to extrapolate to all cases, not to say that to internalize (i.e. learn) all these knowledges can be challenging for a single neural model. 

Acquisition of pronunciation knowledge is a noticeable aspect, and its failure could be the root of TTS errors in end-to-end models like reading “chaos” as “CH-AW-S” or “San Jose” as “S-AA-N JH-OW-S”. It is because not only the data may be unable to cover these edge cases, but also the model may fail to learn such outliers sparsely distributed in the data. These two might be solved by a simple lexicon look-up, but how about “desert” in “get their just desert” versus “lost in a desert”? It can be even more severe: Thousands of Chinese characters are used in Japanese, each has multiple readings depending on the context. Chinese uses more characters that take native speakers years to memorize. Arabic and Hebrew scripts typically omit vowels, leaving the reader to decide the meaning and pronunciation. 
%For example, as for pronunciation knowledge, natural languages rarely have perfectly phonemic orthography. There are full of English words not covered by general pronunciation rules, and logographic scripts like Chinese almost have no such general rules. More difficultly, the cannot be solved simply by a lexicon lookup due to heteronyms or polyphones: a single character in Japanese may have 10 different readings, and in Abjads like Arabic and Hebrew, vowels are often omitted. The exact reading must be inferred based on the context. 
On these languages without phonemic orthography, fully end-to-end TTS models are often less reliable, needing an extra manually-built complicated grapheme-to-phoneme (G2P) pipeline based on a combination of hand-crafted rules, structured G2P-oriented lexicons, and machine learning models \cite{DBLP:journals/csl/YasudaWY21,fong2019comparison,DBLP:conf/interspeech/TaylorR19}. In this way, additional knowledge and labor from language experts are introduced, which take considerable costs on every single language, unscalable to be extended to countless languages in the real world.

Following research on scalability-centered low-resource TTS \cite{DBLP:journals/corr/abs-2103-03541}, we aim at scaling end-to-end TTS to various languages using minimal per-language human efforts. Given the difficulty both to directly internalize all the knowledge from data and to leverage external knowledge through language expertise, we consider it desirable if external knowledge could be \textbf{automatically} extracted and utilized. %seek methods to leverage  general methods to directly extract external knowledge from easy-to-find resources. 
This is inspired by machine reading comprehension (MRC) and question answering (QA): With rich knowledge learned, BERT may answer questions like “The capital city of Italy is [MASK]” by themselves \cite{DBLP:conf/emnlp/PetroniRRLBWM19}. But the answers are often imprecise as it is difficult and inefficient for a neural network to memorize countless definite or “hard” answers (like the capital of Italy) sporadically appeared in training data. Instead, open-domain QA models are more reliable on this task. They do not store such knowledge inside themselves, but develop a “soft” ability to read and understand related external texts (like Wikipedia articles), and then find the correct answer to the question from the article \cite{DBLP:conf/iclr/SeoKFH17,DBLP:conf/acl/ChenFWB17}.

\begin{figure}[t]
% \vskip 0.2in
\centerline{\includegraphics[width=\columnwidth]{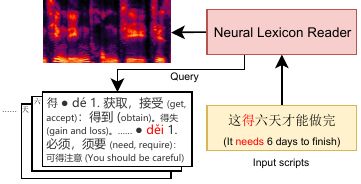}}
\caption{Neural Lexicon Reader under the framework, in which raw lexicon text of each character is used as external knowledge. The lexicon contains readings as well as meanings and usages for each reading. The character “\includegraphics[height=0.8em]{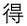}” takes the second reading děi here, since it means “need” in the sentence. English translations are indicated.}
\label{Fig:framework}
% \vskip -0.2in
\end{figure}

TTS resembles QA in that both the hard knowledge (e.g. pronunciations and semantics of each word) and soft ability (e.g. to choose the exact reading) are required.
Therefore, we design a general framework with novel Token2Knowledge attention to automatically extract relevant knowledge from external texts. To exemplify, we propose Neural Lexicon Reader (NLR) shown in Figure~\ref{Fig:framework}, which extracts pronunciations of words in the input script from raw lexicon texts without a G2P pipeline using structured rules hand-crafted on the language. 
%Various information including syntactic, semantic, and phonetic knowledge can be introduced in this way. 
%Particularly, we demonstrate the potential of the framework to leverage phonetic knowledge by applying it to end-to-end TTS without the G2P pipeline.
%The framework may read the lexicon entry and extract hidden representations of the correct pronunciation.
It mimics humans that when a child doesn't know how to read “chaos”, the child will look it up in a dictionary. More, general-purpose dictionaries are available in most languages, hence the framework can be easily extended to new languages. 

Moreover, it is more flexible, controllable, and interpretable: The model is targeted to develop a more general lexicon-reading capability to be used in runtime that is less dependent on the exact content of scripts and lexicons seen during training. Hence even for a trained model, pronunciations can be manipulated and new words can be actively added by updating the lexicon. Empirical studies on Mandarin Chinese show that NLR produces better intelligibility than character-based models, particularly with limited data. It also helps to resolve heteronyms or polyphones, and generalizes to rare and even unseen characters well, which are effectively zero-shot learning. Furthermore, the lexicon-reading capability is less language-specific and can be better transferred to related languages like Cantonese and Japanese with much less data.

%Pronunciation knowledge could be a simple one to handle, and an additional G2P step seems to be more practical. But this exploration may open a door for leveraging external knowledge in a more explicit way, such as semantic and emotional ones.

To conclude, our contributions are three-fold:

1. We propose a general framework to leverage external knowledge in neural TTS using Token2Knowledge attention;

2. We apply the framework to achieve fully end-to-end TTS on non-phonemic scripts without the need of particular language expertise for building G2P;

3. We demonstrate its general lexicon-reading ability using low-resource and cross-language transfer scenarios.

Code, audio samples, and the model using open data are available at \url{mutiann.github.io/papers/nlr}.

\section{Methods}
\label{sec:methods}
\begin{figure}[t]
% \vskip 0.2in
\centerline{\includegraphics[width=\columnwidth]{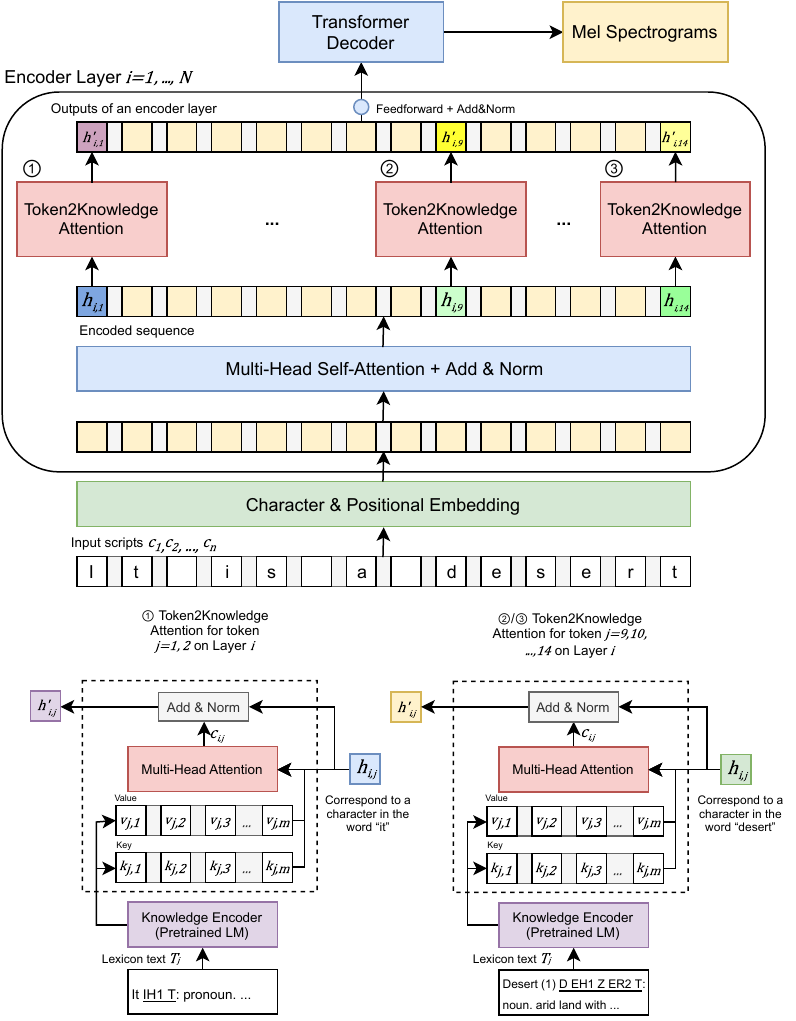}}
\caption{Model architecture, using an English sentence and a dictionary with ARPABET pronunciation notation as an example.  \raisebox{.5pt}{\textcircled{\raisebox{-.9pt} {1}}} is computed upon an element in the encoded sequence that corresponds to “I” in “It”, while \raisebox{.5pt}{\textcircled{\raisebox{-.9pt} {2}}} corresponds to “d” and \raisebox{.5pt}{\textcircled{\raisebox{-.9pt} {3}}} to “t” in “desert”. Therefore \raisebox{.5pt}{\textcircled{\raisebox{-.9pt} {2}}} and \raisebox{.5pt}{\textcircled{\raisebox{-.9pt} {3}}} use the same dictionary entry for “desert” for the Token2Knowledge Attention. The Token2Knowledge Attention modules are identical using shared parameters, but lexicon texts and input elements from the encoded sequence are different.}
\label{Fig:model}
% \vskip -0.2in
\end{figure}

We follow \cite{DBLP:journals/corr/abs-2103-03541} to use a transformer model, but with character inputs and 4 decoder layers, and Token2Knowledge attention introduced in the encoder, as illustrated in Figure~\ref{Fig:model}: For each token $c_j$, possibly a character, its relevant textual knowledge $T_{j,1}, T_{j,2}, ..., T_{j,m}$ (possibly the lexicon entry of the word it belongs to) is passed through a \textbf{knowledge encoder} to create two sequences of hidden representations: \textbf{keys} $k_{j,1}, k_{j,2}, ..., k_{j,m}$ and \textbf{values} $v_{j,1}, v_{j,2}, ..., v_{j,m}$. Then, for the hidden state $h_{i,j}$ at encoder layer $i$, we compute a context vector of 
\begin{equation}
\begin{aligned}
    c_{i,j} = \text{attention}(h_{i,j}, [k_{j,1}, ..., k_{j,m}], 
    [v_{j,1}, ..., v_{j,m}])
\end{aligned}
\end{equation}

by the standard multi-head attention layer that extracts relevant values according to the matchness between $h_{i,j}$ and the key, 
% in which 
%
% \begin{equation}
% \begin{aligned}
%     attention(Q, K, V) = softmax(\frac{QK^T}{\sqrt{d_k})})V
% \end{aligned}
% \end{equation}
followed by $h'_{i,j} = \text{Layernorm}(c_{i,j} + h_{i,j})$. This is similar to the transformer decoder, in which the decoder states perform not only self-attention but also attention with encoder states. The difference is that, in the decoder, all elements in the sequence of decoder states attend to the same sequence of encoder states. While in Token2Knowledge Attention, different elements attend to different encoded external knowledge.

The framework is flexible: the form of knowledge, the knowledge encoder, and the way to select relevant knowledge can be decided according to the scenario. As for NLR to extract pronunciation knowledge in our experiments, raw texts from the corresponding lexicon entry are used as knowledge, and the pretrained multilingual language model XLM-R \cite{DBLP:conf/acl/ConneauKGCWGGOZ20} is used as the knowledge encoder. Final-layer embeddings from XLM-R are used as  \textbf{keys} to capture the semantics of the lexicon texts, so as to match the semantics between input texts and lexicon texts. While to represent pronunciations, exact identities of characters (such as those in the phonemic notations) in the lexicon are valuable to present in attention outputs. Therefore, \textbf{values} are built by a concatenation of final-layer and first-layer embeddings of XLM-R, covering both deep (semantic) and shallow (identity) information. For unknown tokens and punctuation without lexicon entries, keys and values are both zero vectors. Token2Knowledge layers are used in all except the first encoder layers. Lexicon texts are from online dictionary websites. \footnote{~\url{dictionary.goo.ne.jp} for Japanese, \url{humanum.arts.cuhk.edu.hk/Lexis/lexi-can} for Cantonese, and \url{zdic.net} for Mandarin.}

\section{Experiments}
\label{sec:experiments}

We experimented on internal data of multiple East Asian languages, which exemplify the challenges we try to handle, as capturing their pronunciation is difficult, particularly in low-resource cases: There are no general rules, requiring case-by-case memorization; Readings are context-dependent, so a simple lexicon-lookup is not enough. They are also widely spoken but rather diverse, many of them (like various Sinitic dialects) under-resourced, making the scenario realistic. Nevertheless, our method is applicable beyond them, as many other languages such as Arabic and Indic ones share similar challenges.

To evaluate the method's capability to leverage external knowledge when training data fail to cover necessary knowledge, we experimented on simulated low-resource cases with a single speaker Mandarin dataset of 18K samples downsampled to different sizes, without speaker or language embeddings. Besides, to evaluate the transferability of the lexicon-reading capability, we train a base model on a larger Mandarin dataset with 93k samples from 11 speakers, and adapt it to single-speaker Cantonese or Japanese data. The base model uses only speaker embeddings, so language embeddings are added to input embeddings and trained. All data are without word segmentation. Results are evaluated in comparison with baselines under the same setting but without Token2Knowledge layers. Other settings follow the prior work \cite{DBLP:journals/corr/abs-2103-03541}. Waveforms are reconstructed from synthesized mel-spectrograms of a 100-sample held-out set on each language using the Griffin-Lim algorithm, quality from which sufficient for intelligibility evaluation. They are then sent to Azure Speech-to-Text to report objective character error rate (CER). Although the black-box recognizer has errors, it is generally reliable for comparing prevalence of pronunciation errors between systems \cite{DBLP:conf/interspeech/TaylorR21}. A pretrained WaveNet is used to generate waveforms for subjective evaluations.

\subsection{Mandarin experiments}

\begin{table}[t]
\caption{Objective CER(\%) for Mandarin systems}
\label{Table:mandarin}
\vskip 0.1in
\begin{center}
\begin{small}
\begin{sc}
\begin{tabular}{lllll}
\toprule
Dataset size & 18K  & 10K  & 7.5K  & 5K    \\
\midrule
Baseline                    & 4.65 & 9.40  & 18.33 & Fail  \\
NLR                         & 4.82 & 5.86 & \textbf{7.14}  & \textbf{13.64} \\
\midrule
+Word Query & \textbf{4.39} & \textbf{4.71} & 7.28  & 14.78 \\
w/o XLM-R                   & 4.84 & 6.11 & 10.50  & 15.13 \\
\bottomrule
\end{tabular}
\end{sc}
\end{small}
\end{center}
\vskip -0.2in
\end{table}

\begin{table}[t]
\caption{Subjective error rate (\%) on different test sets}
\vskip 0.1in
\label{Table:domain}
\begin{center}
\begin{small}
\begin{sc}
\begin{tabular}{lllll}
\toprule
& \multicolumn{2}{c}{Rare} & \multicolumn{2}{c}{Heteronyms} \\ 
\cmidrule(l{2pt}r{2pt}){2-3} \cmidrule(l{2pt}r{2pt}){4-5}
Dataset size & 18K & 10K & 18K & 10K \\ 
\midrule
Baseline & 8.0 & 62.0 & 75.5 & 80.9 \\
NLR & 2.0 & 4.0 & 72.6 & 76.4 \\

\midrule

& \multicolumn{4}{c}{General} \\
\cmidrule(l{2pt}r{2pt}){2-5}

Dataset size & 18K & 10K & 7.5K & 5K \\ 
\midrule

Baseline & 0.9 & 5.4 & 10.9 & Fail \\
NLR & 0.3 & 1.9 & 4.1 & 8.5 \\

\bottomrule
\end{tabular}
\end{sc}
\end{small}
\end{center}

\vskip -0.2in
\end{table}

Objective CERs are given in Table~\ref{Table:mandarin}: when using the complete 18K-sample data, baseline and NLR models produce similar results, with CER both close to the ground truth one of 4.35\%. But with fewer samples, NLR models are more resistant to reduced data and constantly outperform the baseline (which fails to converge at 5K data), showing that NLR successfully leverages pronunciation knowledge from lexicon texts when knowledge from data is insufficient. +Word Query and w/o XLM-R models will be discussed in Section~\ref{Sec:ablation}. 

Beside the general cases, we create test sets focused on scenarios with special challenges. First, for the capability to generalize to rare and even unseen characters (i.e. few-shot or zero-shot cases), we create a 100-utterance \textsc{Rare} test set. Each of the utterances contains an uncommon character. They only appear once in the full 18K data, and 50 of them unseen in the 10K subset. To avoid interference, other used characters are simple to handle. 18K and 10K-data baseline and NLR models with the lowest objective CERs are used to synthesize the speeches to be inspected by a language expert. As shown in Table~\ref{Table:domain}, both NLR models give correct pronunciations in most cases regardless of dataset sizes. Baselines show more errors to internalize the knowledge that is sparse in the training data, not to say the 10K case when such knowledge is unavailable. It also shows the extensibility of the framework that new knowledge (like pronunciations of unseen characters) can be directly introduced to a trained model. An experiment further supports this: we train a model with 10K data but remove lexicon entries for characters unseen in the 2K subset to simulate an incomplete lexicon, and the best CER can only reach 6.38\%. But if the full lexicon is additionally given on evaluation, the CER reaches 5.91\%, indicating that the model leverages and generalizes to new lexicon entries unseen during training.
%This model has a 6.15\% objective CER, similar to the standard NLR model, though multiple unseen lexicon entries appear in the test from the model's perspective.
%The result shows that external knowledge is leveraged by NLR.

Similarly, we create a rather challenging \textsc{Heteronym} test set. We choose 115 characters with multiple readings. For each character, utterances are created using each reading except the most frequent one in training, hence a baseline that simply uses the most frequent reading will have 100\% error rate. 
Total 292 utterances are used, and the error rates averaged by each reading of each character are reported. As shown in Table~\ref{Table:domain}, NLR obtain improved ability regarding resolving heteronyms, showing an increased “soft” ability to decide the pronunciation according to the context, although the test remains challenging.

Furthermore, we evaluate subjective CERs on synthetic speech used to report Table~\ref{Table:mandarin}. Given in Table~\ref{Table:domain} as the \textsc{General} CER, NLR constantly outperforms the baseline. We also find that low-resource cases errors in NLR are more due to misalignments, not mispronunciations. Besides, the curve is consistent with the objective metrics, showing the reliability to compare intelligibility with objective CER.

\begin{table}[t]
\caption{CER(\%) for low-resource adaptation to a different language, with different dataset size for each column}
\label{Table:adapt}
\begin{center}
\begin{small}
\begin{sc}
\begin{tabular}{llllll}
\toprule
Cantonese       & 2K    & 1K    & 750   & 500   & 250   \\
\midrule
Baseline & 12.32 & 14.89 & 17.64 & 22.27 & 35.08 \\
NLR      & 8.79  & 10.04 & 10.26 & 10.73 & 12.85 \\
\midrule
Japanese       & 5K    & 3K    & 2K    & 1K    & 750   \\
\midrule
Baseline & 12.46 & 15.99 & 18.66 & 26.85 & 33.76 \\
NLR      & 10.50 & 12.48 & 13.95 & 19.29 & 21.81 \\
\bottomrule
\end{tabular}
\end{sc}
\end{small}
\end{center}
\vskip -0.2in
\end{table}

\begin{figure*}[t]
% \vskip 0.2in
\centerline{\includegraphics[width=\textwidth]{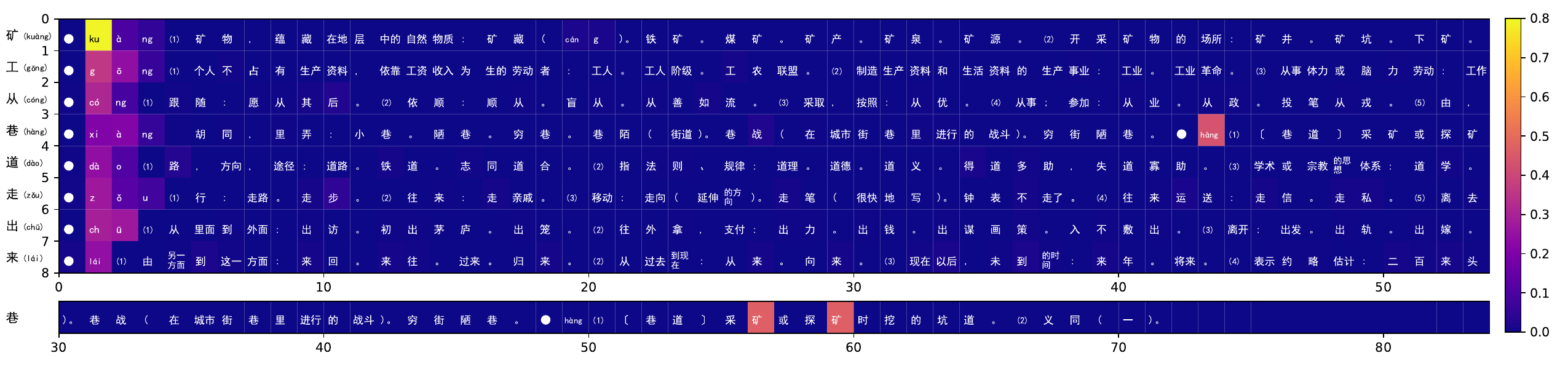}}
\caption{An alignment heatmap from Token2Knowledge attention with input tokens labelled on y-axis, and corresponding lexicon texts in the grid along x-axis, plus the row of the heatmap for “\includegraphics[height=0.8em]{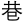}” in another head. Lexicon texts that are too long are omitted.}
\label{Fig:align}
% \vskip -0.2in
\end{figure*}

\subsection{Cross-language transfer learning}

Cantonese is a Chinese variant with systematic phonetic differences from Mandarin, making them mutually unintelligible. However, as shown by the results in Table~\ref{Table:adapt}, with only 2K Cantonese samples the CER of an adapted NLR is close to the ground truth (8.45\%), and CERs increase slowly when using fewer samples. Japanese is more difficult: unlike Chinese, almost all characters have multiple readings. The native \textit{kun-yomi} readings are not of Chinese origin, and are often multisyllabic which adds to the modeling difficulty. As a result, the CER gap between the 5K model and the ground truth (7.20\%) is large. Nevertheless, the results are constantly improved with NLR, showing that the lexicon-reading capability is not fully language-dependent, and models trained on one language can be transferred to a distant language with fewer data.

\subsection{Comparative study}
\label{Sec:ablation}
 We attempt to perform word segmentation on the texts and query the lexicon on a word-level granularity. In this way, the intonation and heteronym resolution might be easier. As shown by the \textsc{+Word Query} model in Table~\ref{Table:mandarin}, the performance is slightly improved with relatively rich data. However, it harms the performance with few data, possibly because representing readings of words but not monosyllabic characters is more difficult. Hence, word segmentation is generally not essential for NLR. We also tried to remove pretrained XLM-R, but use the same set of character embeddings to encode both inputs and lexicon text values. To capture higher-level semantic information, an additional transformer encoder layer is applied upon the embeddings to produce the keys. As shown in \textsc{w/o XLM-R}, limited regression can be observed in the results, particularly with fewer data, which shows that our method might be also applied to languages without such a pretrained model.

\subsection{Case study}

Figure~\ref{Fig:align} shows an alignment heatmap from the 18K-data Mandarin model synthesizing the given utterance. Pinyins of the correct pronunciations have a high alignment weight, indicating that the model extracts the correct pinyin from the lexicon entry, while irrelevant pinyins (like column 19 “cáng” in row 0 “\includegraphics[height=0.8em]{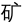}”) are ignored. While the character “\includegraphics[height=0.8em]{symbols/xiang.pdf}” is a heteronym, and under this context the character should be pronounced with an infrequent reading “hàng” which is only used in mining terms. The model produces the correct reading, which can be interpreted by the larger alignment weight assigned to “hàng” instead of “xiàng” in the lexicon entry. This is possibly because another head of attention for “\includegraphics[height=0.8em]{symbols/xiang.pdf}” extracts context vectors from tokens “\includegraphics[height=0.8em]{symbols/kuang.pdf}” (“mine”) in the lexicon entry, as shown in the lower half of the figure. More, the pronunciation can be directly manipulated by changing lexicon texts fed to the model for each character, such as by adding, modifying, or removing pronunciations in the text, as given in our audio samples, showing better controllability than ordinary end-to-end TTS or G2P models.

\begin{table}[t]
\caption{Mean opinion score of different voices, with 95\% confidence interval}
\label{Table:mos}
\begin{center}
\begin{small}
\begin{sc}
\begin{tabular}{lll}
\toprule
         Dataset size & 18K & 10K \\ 
         \midrule
        Baseline & 3.84$\pm$0.06 & 3.65$\pm$0.07 \\ 
        NLR & \textbf{3.85$\pm$0.06} & \textbf{3.80$\pm$0.06} \\ 
         \midrule
         Ground Truth & \multicolumn{2}{c}{3.98$\pm$0.07} \\
\bottomrule
\end{tabular}
\end{sc}
\end{small}
\end{center}

\vskip -0.2in
\end{table}

\subsection{Naturalness evaluation}
 To evaluate the naturalness, we randomly sampled 50 utterances from the Mandarin held-out set. Mean opinion scores (MOS) are given by 20 judges per utterance. As shown in Table~\ref{Table:mos}, our NLR model shows good naturalness, on par with the baseline with rich resources and better under lower resources.

\section{Related work}
\label{sec:related}
Our work is closely related to Byte2Speech \cite{DBLP:journals/corr/abs-2103-03541}, as we follow their transformer-based model and scalability-centered goal, while their performances on non-phonemic scripts are unsatisfying. The framework is inspired by models that extract textual knowledge \cite{DBLP:conf/iclr/SeoKFH17,DBLP:conf/acl/ChenFWB17}, possibly from lexicon, for commonsense reasoning \cite{DBLP:conf/emnlp/BauerWB18}, QA \cite{DBLP:conf/acl/WangJ19}, and Chinese sequence labelling \cite{DBLP:conf/acl/LiuFZX20}. Lexicon texts have been used in word sense disambiguation \cite{DBLP:conf/emnlp/HuangSQH19} and machine translation (MT) \cite{DBLP:journals/corr/abs-2010-05997}, while we apply the idea to TTS. The particular task we apply to is closely related to G2P and polyphone disambiguation with rich prior work, particularly on Chinese, Japanese, and Arabic \cite{DBLP:conf/interspeech/NiSK18,DBLP:conf/icassp/YuNSLSCMK20,DBLP:conf/interspeech/ZhangPL20,DBLP:conf/wanlp/AlqahtaniAD19}, possibly enhanced by explicit rules \cite{DBLP:conf/lrec/GormanMN18} or implicit external knowledge \cite{DBLP:conf/interspeech/DaiWKW0S0M19}, and lexicons have been used during training \cite{DBLP:conf/interspeech/ZouD019,shi21d_interspeech}. While the flexibility of using external knowledge has been discussed in MT \cite{DBLP:conf/emnlp/FengZZWA17} and G2P \cite{DBLP:conf/interspeech/BruguierBS18}, though knowledge from raw lexicon texts were not leveraged. Representations from pretrained language models introduce additional knowledge and enhance TTS and G2P performances \cite{DBLP:conf/interspeech/HayashiWTTTL19,DBLP:conf/icassp/XiaoHMS20,pngbert,nicolis21_ssw,DBLP:conf/icassp/HidaHKTSK22}, but in most cases, the training data contain limited phonemic information, making these models less useful for capturing pronunciations. Also, knowledge from the neural models is lack of explicit representation and control.

\section{Conclusion and future work}
\label{sec:conclusion}
We demonstrate that knowledge can be extracted from external raw texts in end-to-end TTS models, with the application of Neural Lexicon Reader that obtains pronunciations from lexicon entries. Future work may include improving the extraction capability and applying it to other types of knowledge such as semantic meanings to improve speech prosody and naturalness.

\bibliographystyle{IEEEtran}

\bibliography{mybib}

\begin{thebibliography}{10}
\providecommand{\url}[1]{#1}
\csname url@samestyle\endcsname
\providecommand{\newblock}{\relax}
\providecommand{\bibinfo}[2]{#2}
\providecommand{\BIBentrySTDinterwordspacing}{\spaceskip=0pt\relax}
\providecommand{\BIBentryALTinterwordstretchfactor}{4}
\providecommand{\BIBentryALTinterwordspacing}{\spaceskip=\fontdimen2\font plus
\BIBentryALTinterwordstretchfactor\fontdimen3\font minus
  \fontdimen4\font\relax}
\providecommand{\BIBforeignlanguage}[2]{{%
\expandafter\ifx\csname l@#1\endcsname\relax
\typeout{** WARNING: IEEEtran.bst: No hyphenation pattern has been}%
\typeout{** loaded for the language `#1'. Using the pattern for}%
\typeout{** the default language instead.}%
\else
\language=\csname l@#1\endcsname
\fi
#2}}
\providecommand{\BIBdecl}{\relax}
\BIBdecl

\bibitem{DBLP:conf/icassp/ShenPWSJYCZWRSA18}
J.~Shen, R.~Pang, R.~J. Weiss, M.~Schuster, N.~Jaitly, Z.~Yang, Z.~Chen,
  Y.~Zhang, Y.~Wang, R.~Ryan, R.~A. Saurous, Y.~Agiomyrgiannakis, and Y.~Wu,
  ``Natural {TTS} synthesis by conditioning wavenet on {MEL} spectrogram
  predictions,'' in \emph{Proc. ICASSP 2018}.\hskip 1em plus 0.5em minus
  0.4em\relax {IEEE}, 2018, pp. 4779--4783.

\bibitem{DBLP:journals/csl/YasudaWY21}
Y.~Yasuda, X.~Wang, and J.~Yamagishi, ``Investigation of learning abilities on
  linguistic features in sequence-to-sequence text-to-speech synthesis,''
  \emph{Comput. Speech Lang.}, vol.~67, p. 101183, 2021.

\bibitem{fong2019comparison}
J.~Fong, J.~Taylor, K.~Richmond, and S.~King, ``A comparison between letters
  and phones as input to sequence-to-sequence models for speech synthesis,'' in
  \emph{The 10th ISCA Speech Synthesis Workshop}.\hskip 1em plus 0.5em minus
  0.4em\relax ISCA, 2019, pp. 223--227.

\bibitem{DBLP:conf/interspeech/TaylorR19}
J.~Taylor and K.~Richmond, ``Analysis of pronunciation learning in end-to-end
  speech synthesis,'' in \emph{Proc. Interspeech 2019}.\hskip 1em plus 0.5em
  minus 0.4em\relax {ISCA}, 2019, pp. 2070--2074.

\bibitem{DBLP:journals/corr/abs-2103-03541}
M.~He, J.~Yang, L.~He, and F.~K. Soong, ``Multilingual {B}yte2{S}peech models
  for scalable low-resource speech synthesis,'' \emph{CoRR}, vol.
  abs/2103.03541, 2021.

\bibitem{DBLP:conf/emnlp/PetroniRRLBWM19}
F.~Petroni, T.~Rockt{\"{a}}schel, S.~Riedel, P.~S.~H. Lewis, A.~Bakhtin, Y.~Wu,
  and A.~H. Miller, ``Language models as knowledge bases?'' in \emph{Proc.
  EMNLP/IJCNLP 2019}.\hskip 1em plus 0.5em minus 0.4em\relax ACL, 2019, pp.
  2463--2473.

\bibitem{DBLP:conf/iclr/SeoKFH17}
M.~J. Seo, A.~Kembhavi, A.~Farhadi, and H.~Hajishirzi, ``Bidirectional
  attention flow for machine comprehension,'' in \emph{Proc. ICLR 2017}.\hskip
  1em plus 0.5em minus 0.4em\relax OpenReview.net, 2017.

\bibitem{DBLP:conf/acl/ChenFWB17}
D.~Chen, A.~Fisch, J.~Weston, and A.~Bordes, ``Reading {W}ikipedia to answer
  open-domain questions,'' in \emph{Proc. ACL 2017}.\hskip 1em plus 0.5em minus
  0.4em\relax ACL, 2017, pp. 1870--1879.

\bibitem{DBLP:conf/acl/ConneauKGCWGGOZ20}
A.~Conneau, K.~Khandelwal, N.~Goyal, V.~Chaudhary, G.~Wenzek, F.~Guzm{\'{a}}n,
  E.~Grave, M.~Ott, L.~Zettlemoyer, and V.~Stoyanov, ``Unsupervised
  cross-lingual representation learning at scale,'' in \emph{Proc. ACL
  2020}.\hskip 1em plus 0.5em minus 0.4em\relax ACL, 2020, pp. 8440--8451.

\bibitem{DBLP:conf/interspeech/TaylorR21}
J.~Taylor and K.~Richmond, ``Confidence intervals for {ASR}-based {TTS}
  evaluation,'' in \emph{Proc. Interspeech 2021}.\hskip 1em plus 0.5em minus
  0.4em\relax {ISCA}, 2021, pp. 2791--2795.

\bibitem{DBLP:conf/emnlp/BauerWB18}
L.~Bauer, Y.~Wang, and M.~Bansal, ``Commonsense for generative multi-hop
  question answering tasks,'' in \emph{Proc. EMNLP 2018}.\hskip 1em plus 0.5em
  minus 0.4em\relax ACL, 2018, pp. 4220--4230.

\bibitem{DBLP:conf/acl/WangJ19}
C.~Wang and H.~Jiang, ``Explicit utilization of general knowledge in machine
  reading comprehension,'' in \emph{Proc. ACL 2019}.\hskip 1em plus 0.5em minus
  0.4em\relax ACL, 2019, pp. 2263--2272.

\bibitem{DBLP:conf/acl/LiuFZX20}
W.~Liu, X.~Fu, Y.~Zhang, and W.~Xiao, ``Lexicon enhanced {C}hinese sequence
  labeling using {BERT} adapter,'' in \emph{Proc. ACL/IJCNLP 2021}.\hskip 1em
  plus 0.5em minus 0.4em\relax ACL, 2021, pp. 5847--5858.

\bibitem{DBLP:conf/emnlp/HuangSQH19}
L.~Huang, C.~Sun, X.~Qiu, and X.~Huang, ``Gloss{BERT}: {BERT} for word sense
  disambiguation with gloss knowledge,'' in \emph{Proc. EMNLP-IJCNLP
  2019}.\hskip 1em plus 0.5em minus 0.4em\relax ACL, 2019, pp. 3507--3512.

\bibitem{DBLP:journals/corr/abs-2010-05997}
X.~J. Zhong and D.~Chiang, ``Look it up: Bilingual and monolingual dictionaries
  improve neural machine translation,'' \emph{CoRR}, vol. abs/2010.05997, 2020.

\bibitem{DBLP:conf/interspeech/NiSK18}
J.~Ni, Y.~Shiga, and H.~Kawai, ``Multilingual grapheme-to-phoneme conversion
  with global character vectors,'' in \emph{Proc. Interspeech 2018}.\hskip 1em
  plus 0.5em minus 0.4em\relax {ISCA}, 2018, pp. 2823--2827.

\bibitem{DBLP:conf/icassp/YuNSLSCMK20}
M.~Yu, H.~D. Nguyen, A.~Sokolov, J.~Lepird, K.~M. Sathyendra, S.~Choudhary,
  A.~Mouchtaris, and S.~Kunzmann, ``Multilingual grapheme-to-phoneme conversion
  with byte representation,'' in \emph{Proc. ICASSP 2020}.\hskip 1em plus 0.5em
  minus 0.4em\relax {IEEE}, 2020, pp. 8234--8238.

\bibitem{DBLP:conf/interspeech/ZhangPL20}
H.~Zhang, H.~Pan, and X.~Li, ``A mask-based model for {Mandarin Chinese}
  polyphone disambiguation,'' in \emph{Proc. Interspeech 2020}.\hskip 1em plus
  0.5em minus 0.4em\relax {ISCA}, 2020, pp. 1728--1732.

\bibitem{DBLP:conf/wanlp/AlqahtaniAD19}
S.~Alqahtani, H.~Aldarmaki, and M.~T. Diab, ``Homograph disambiguation through
  selective diacritic restoration,'' in \emph{Proc. Fourth Arabic Natural
  Language Processing Workshop, WANLP@ACL 2019}.\hskip 1em plus 0.5em minus
  0.4em\relax ACL, 2019, pp. 49--59.

\bibitem{DBLP:conf/lrec/GormanMN18}
K.~Gorman, G.~Mazovetskiy, and V.~Nikolaev, ``Improving homograph
  disambiguation with supervised machine learning,'' in \emph{Proc. {LREC}
  2018}.\hskip 1em plus 0.5em minus 0.4em\relax European Language Resources
  Association {(ELRA)}, 2018.

\bibitem{DBLP:conf/interspeech/DaiWKW0S0M19}
D.~Dai, Z.~Wu, S.~Kang, X.~Wu, J.~Jia, D.~Su, D.~Yu, and H.~Meng,
  ``Disambiguation of {C}hinese polyphones in an end-to-end framework with
  semantic features extracted by pre-trained {BERT},'' in \emph{Proc.
  Interspeech 2019}.\hskip 1em plus 0.5em minus 0.4em\relax {ISCA}, 2019, pp.
  2090--2094.

\bibitem{DBLP:conf/interspeech/ZouD019}
Y.~Zou, L.~Dong, and B.~Xu, ``Boosting character-based {C}hinese speech
  synthesis via multi-task learning and dictionary tutoring,'' in \emph{Proc.
  Interspeech 2019}.\hskip 1em plus 0.5em minus 0.4em\relax {ISCA}, 2019, pp.
  2055--2059.

\bibitem{shi21d_interspeech}
Y.~Shi, C.~Wang, Y.~Chen, and B.~Wang, ``Polyphone disambiguation in {M}andarin
  {C}hinese with semi-supervised learning,'' in \emph{Proc. Interspeech
  2021}.\hskip 1em plus 0.5em minus 0.4em\relax {ISCA}, 2021, pp. 4109--4113.

\bibitem{DBLP:conf/emnlp/FengZZWA17}
Y.~Feng, S.~Zhang, A.~Zhang, D.~Wang, and A.~Abel, ``Memory-augmented neural
  machine translation,'' in \emph{Proc. EMNLP 2017}.\hskip 1em plus 0.5em minus
  0.4em\relax ACL, 2017, pp. 1390--1399.

\bibitem{DBLP:conf/interspeech/BruguierBS18}
A.~Bruguier, A.~Bakhtin, and D.~Sharma, ``Dictionary augmented
  sequence-to-sequence neural network for grapheme to phoneme prediction,'' in
  \emph{Proc. Interspeech 2018}.\hskip 1em plus 0.5em minus 0.4em\relax {ISCA},
  2018, pp. 3733--3737.

\bibitem{DBLP:conf/interspeech/HayashiWTTTL19}
T.~Hayashi, S.~Watanabe, T.~Toda, K.~Takeda, S.~Toshniwal, and K.~Livescu,
  ``Pre-trained text embeddings for enhanced text-to-speech synthesis,'' in
  \emph{Proc. Interspeech 2019}.\hskip 1em plus 0.5em minus 0.4em\relax {ISCA},
  2019, pp. 4430--4434.

\bibitem{DBLP:conf/icassp/XiaoHMS20}
Y.~Xiao, L.~He, H.~Ming, and F.~K. Soong, ``Improving prosody with linguistic
  and {BERT} derived features in multi-speaker based {M}andarin {C}hinese
  neural {TTS},'' in \emph{Proc. ICASSP 2020}.\hskip 1em plus 0.5em minus
  0.4em\relax {IEEE}, 2020, pp. 6704--6708.

\bibitem{pngbert}
Y.~Jia, H.~Zen, J.~Shen, Y.~Zhang, and Y.~Wu, ``{PnG BERT: A}ugmented {BERT} on
  phonemes and graphemes for neural {TTS},'' in \emph{Proc. Interspeech
  2021}.\hskip 1em plus 0.5em minus 0.4em\relax {ISCA}, 2021, pp. 151--155.

\bibitem{nicolis21_ssw}
M.~Nicolis and V.~Klimkov, ``{Homograph disambiguation with contextual word
  embeddings for {TTS} systems},'' in \emph{The 11th ISCA Speech Synthesis
  Workshop}.\hskip 1em plus 0.5em minus 0.4em\relax ISCA, 2021, pp. 222--226.

\bibitem{DBLP:conf/icassp/HidaHKTSK22}
R.~Hida, M.~Hamada, C.~Kamada, E.~Tsunoo, T.~Sekiya, and T.~Kumakura,
  ``Polyphone disambiguation and accent prediction using pre-trained language
  models in {J}apanese {TTS} front-end,'' in \emph{Proc. ICASSP 2022}.\hskip
  1em plus 0.5em minus 0.4em\relax {IEEE}, 2022, pp. 7132--7136.

\end{thebibliography}

% \begin{thebibliography}{9}
% \bibitem[1]{Davis80-COP}
%   S.\ B.\ Davis and P.\ Mermelstein,
%   ``Comparison of parametric representation for monosyllabic word recognition in continuously spoken sentences,''
%   \textit{IEEE Transactions on Acoustics, Speech and Signal Processing}, vol.~28, no.~4, pp.~357--366, 1980.
% \bibitem[2]{Rabiner89-ATO}
%   L.\ R.\ Rabiner,
%   ``A tutorial on hidden Markov models and selected applications in speech recognition,''
%   \textit{Proceedings of the IEEE}, vol.~77, no.~2, pp.~257-286, 1989.
% \bibitem[3]{Hastie09-TEO}
%   T.\ Hastie, R.\ Tibshirani, and J.\ Friedman,
%   \textit{The Elements of Statistical Learning -- Data Mining, Inference, and Prediction}.
%   New York: Springer, 2009.
% \bibitem[4]{YourName17-XXX}
%   F.\ Lastname1, F.\ Lastname2, and F.\ Lastname3,
%   ``Title of your INTERSPEECH 2022 publication,''
%   in \textit{Interspeech 2022 -- 23\textsuperscript{rd} Annual Conference of the International Speech Communication Association, September 18-22, Incheon, Korea, Proceedings, Proceedings}, 2022, pp.~100--104.
% \end{thebibliography}

\end{document}